\documentclass[iop]{emulateapj}
\usepackage{apjfonts}
\usepackage{hyperref}
\usepackage{natbib,times}

\def\B{{\rm B}}
\def\d{{d}}
\def\H{{\rm H}}
\def\mw{{\rm mw}}
\def\sl{{\rm sl}}

\begin{document}


\shorttitle{Characterizing Galaxy Clusters with Gravitational Potential}
\shortauthors{Lau}

\title{Characterizing Galaxy Clusters with Gravitational Potential}

\author{
Erwin T. Lau
}

\affil{Department of Astronomy \& Astrophysics, The University of Chicago, 5640 South Ellis Ave., Chicago, IL 60637, U.S.A; ethlau@oddjob.uchicago.edu\\
and\\
Shanghai Astronomical Observatory, Chinese Academy of Sciences, 80 Nandan Road, Shanghai 200030, China.}
   

\begin{abstract}
We propose a simple estimator for the gravitational potential of cluster-size halos using the 
temperature and density profiles of the intracluster gas 
based on the assumptions of hydrostatic equilibrium and spherical symmetry. 
Using high resolution cosmological simulations of galaxy clusters, 
we show that the scaling relation between this estimator and the gravitational 
potential has a small intrinsic scatter of $\sim 8\%-15\%$, and it is
insensitive to baryon physics outside the cluster core. The slope and the normalization
of the scaling relation vary weakly with redshift, and they are relatively independent of the choice 
of radial range used and the dynamical states of the clusters. 
The results presented here provide a possible way for using the cluster potential function 
as an alternative to the cluster mass function in constraining cosmology using galaxy clusters. 
\\
\end{abstract}


\section{Introduction}
\label{sec:intro}

Clusters of galaxies are useful probes of cosmology. 
The evolution of their abundance probes the growth of structure and therefore 
provides constraints on cosmological parameters \citep[e.g., see][for reviews]{voit05, allen_etal11}. 
Clusters are readily observable in multiple wavelengths. In addition to the optical, 
the regime where galaxy clusters are first observed,  hot gas in the deep gravitational wells of clusters shines in X-ray. 
In the sub-millimeter regime, galaxy clusters appear as distortions in the cosmic microwave background (CMB) 
as the hot electrons in the intracluster medium (ICM) scatter off CMB photons through inverse Compton scattering 
known as the Sunyaev--Zeldovich (SZ) effect \citep{sunyaev_zeldovich70, carlstrom_etal02}.  
In recent years, cosmological constraints have been yielded using cluster counts alone 
in optical surveys \citep[e.g.,][]{bahcall_etal03,gladders_etal07,rozo_etal10}, in X-ray \citep[e.g.,][]{vikhlinin_etal09,mantz_etal08,mantz_etal10}, 
and in SZ \citep[e.g.,][]{vanderlinde_etal10, sehgal_etal10}. Cluster abundances and
clustering also can also give us constraints on primordial non-gaussianity \citep[e.g.,][]{cunha_etal10} and
models of gravity \citep[e.g.,][]{schmidt_etal09,rapetti_etal10} 

Ideally the simplest way to constrain cosmology is to measure 
the abundance of observable signals from clusters and constrain cosmology directly from them,
e.g., constraining cosmology through direct comparison of the cluster temperature function 
predicted from theory against observations. 
However, current theoretical models are unable to model clusters by their observables to the
precision necessary for constraining cosmological parameters down to percent level.  
On the other hand, theory does provide accurate and precise predictions of the
cluster mass function \citep[e.g.,][]{jenkins_etal01, warren_etal06, tinker_etal08}.  
The current paradigm of cluster cosmology therefore
relies on estimating the cluster mass function from cluster observables 
via well-calibrated observable-mass relations. 

Instead of focusing on cluster mass, an alternative is to characterize clusters by
their gravitational potential, a quantity which is also well predicted by theory. 
Gravitational potential has potentially three advantages over mass. 
First, observable quantities, such as gas temperature and gravitational lensing signals, 
depend on the gravitational potential directly and only indirectly on mass. 
Therefore gravitational potential provides a more direct connection to cluster observables. 
Second, the shape of gravitational potential is more spherical than the matter distribution, and this may reduce the scatter 
in scaling relation due to the non-sphericity of matter distribution. Third, different definitions of halo mass
exists, and they are not necessarily tied to the observable properties of cluster. 
For example, friends-of-friends (FoF) algorithm \citep[e.g.,][]{einasto_etal84,davis_etal85}, frequently used in
defining halos in simulations, is difficult to implement in observations.  
Also, FoF halo in general has irregular shape \citep[e.g.,][]{lukic_etal09} 
and this poses another difficulty when relating to observations, 
where in most cases measurements are made within some circular annulus. 
Another way of defining a halo is through spherical overdensity, 
where average overdensity inside the halo is some value times the mean background density
or the critical density of the universe. While this definition is more 
observationally-oriented, the halo mass defined this way artificially evolves with redshift 
as the mean background density and the critical density both change with redshift. 

Efforts have been recently made both theoretically and observationally in 
characterizing clusters with their gravitational potential.  
\citet{angrick_bartelmann09} derived the abundance of cluster potential minima
using a Gaussian random field formalism \citep{bardeen_etal86}, and was used to predict 
cluster temperature function using a simple spherical-collapse model to model the nonlinear
evolution of the cluster potential. On the observational side, 
\cite{churazov_etal08, churazov_etal10} measured profiles of the gravitational potential 
of elliptical galaxies using both X-ray observations and stellar kinematics to constrain
non-thermal pressure in those galaxies. It should then be straightforward, at least in principle, 
to extend these measurements to group and cluster-size systems. 

In this paper, we show that the cluster potential can be measured reliably via observables, and that
the cluster potential can serve as the defining quantity of cluster.  
We use simulations to estimate the gravitational potential of galaxy clusters by
constructing a simple potential estimator based on intracluster gas profiles, motivated by the observational
results of \cite{churazov_etal08, churazov_etal10}. 
In Section~\ref{sec:sim} we describe the simulations used in the paper. 
In Section~\ref{sec:methods} we describe our estimator of the cluster potential.  
In Section~\ref{sec:results} we show the results of the scaling relations of this potential estimator and the gravitational potential. 
In Section~\ref{sec:summary} we give our summary and discussion. 

\section{Simulations}
\label{sec:sim}

The simulation data we used are identical to those in \citet{nagai_etal07a, nagai_etal07b}, 
where 16 cluster-sized systems are simulated using the Adaptive Refinement Tree
$N$-body$+$gas-dynamics code \citep{kra99,kra02}, which is 
an Eulerian code that uses adaptive refinement in space and time, and non-adaptive
refinement in mass \citep{klypin_etal01} to achieve the dynamic
ranges to resolve the cores of halos formed in self-consistent
cosmological simulations. The simulations assume a flat {$\Lambda$}CDM model: 
$\Omega_{\rm m}=1-\Omega_{\Lambda}=0.3$, $\Omega_{\rm b}=0.04286$,
$h=0.7$ and $\sigma_8=0.9$,
\footnote{Note that this value of $\sigma_8 = 0.9$ is higher than the current
estimate of $\sigma_8=0.8$. We do not expect the results and conclusion of the paper are going to change significantly. 
Nevertheless, a likely difference for a lower $\sigma_8$ would be the later cluster formation time, leading to less dynamically relaxed clusters, potentially increasing the scatter of our potential estimator. } 
where the Hubble constant is defined as
$100h{\ \rm km\ s^{-1}\ Mpc^{-1}}$, and $\sigma_8$ is the mass variance 
within spheres of radius $8\,h^{-1}$~Mpc. The simulations were run using a 
uniform $128^3$ grid with eight levels of mesh refinement.  The box size for
CL101--CL107 is $120\,h^{-1}$~Mpc comoving on a side and is 
$80\,h^{-1}$~Mpc comoving for CL3--CL24. This corresponds to peak spatial
resolution of $\approx 7\,h^{-1}$~kpc and $5\,h^{-1}$~kpc for the two box sizes respectively. 
Only the inner regions $\sim 3$--$10\,h^{-1}$~Mpc surrounding the cluster center were adaptively refined. 
The dark matter (DM) particle mass in the region around each cluster was $m_{\rm p}
\simeq 9.1\times 10^{8}\,h^{-1}\, {M_{\odot}}$ for CL101--107 and
$ m_{\rm p} \simeq 2.7\times 10^{8}\,h^{-1}\,{M_{\odot}}$ for
CL3--24, while other regions were simulated with a bottom mass
resolution.

In Table~\ref{tab:sim} we report $r_{500c}$ (the radius of the cluster within which its average density is
500 times the critical density), $M_{500c}$ (the mass within $r_{500c}$), 
and our classification of the dynamical state of the cluster based on the
morphology of their mock X-ray images. Details of the classification can be found in
\citet{nagai_etal07b}. The cluster center is defined as the location of the potential minimum.

We assess the effects of dissipative gas physics on the relation between the cluster potential and
gas properties by comparing two sets of clusters simulated with the same initial conditions
but with different prescription of gas physics. In the first set, the gas follows simple physics without
any radiative cooling or star formation.  We refer this set of clusters as non-radiative (NR) clusters. 
In the second set, metallicity-dependent radiative cooling, star formation, supernova feedback 
and a uniform UV background were added. We refer this set of clusters as cooling+star formation (CSF) clusters.
For a detailed description of the gas physics implemented in the CSF clusters, please see \citet{nagai_etal07a}.

\begin{table}[t]
\begin{center}
\caption{Properties of the Simulated Clusters at $z=0$}\label{tab:sim}
\begin{tabular}{l c c c c  }
\hline
\hline
Cluster ID\hspace*{5mm} & 
{$M_{500c}$} & 
{$r_{500c}$} & 
Relaxed (1) \\
& {(10$^{14}$ $h^{-1} {M_{\odot}}$)}
& {($h^{-1}$ Mpc)} 
& /Unrelaxed (0) \\
\hline
CL101 \dotfill & 9.02 & 1.16 & 0 \\
CL102 \dotfill & 5.45 & 0.98 & 0 \\
CL103 \dotfill & 5.70 & 0.99 & 0 \\
CL104 \dotfill & 5.40 & 0.98 & 1 \\
CL105 \dotfill & 4.86 & 0.94 & 0 \\
CL106 \dotfill & 3.47 & 0.84 & 0 \\
CL107 \dotfill & 2.57 & 0.76 & 1 \\
CL3 \dotfill & 2.09 & 0.71 & 1 \\
CL5 \dotfill & 1.31 & 0.61 & 1 \\
CL6 \dotfill & 1.68 & 0.66 & 0 \\
CL7 \dotfill & 1.42 & 0.63 & 1 \\
CL9  \dotfill & 0.83 & 0.52 & 0 \\
CL10 \dotfill & 0.67 & 0.49 & 1 \\
CL11 \dotfill & 0.90 & 0.54 & 0 \\
CL14 \dotfill & 0.77 & 0.51 & 1 \\
CL24 \dotfill & 0.35 & 0.39 & 0 \\
\hline
\end{tabular}
\end{center}
\end{table}

\section{Estimator of the Gravitational Potential}
\label{sec:methods}


Following \cite{churazov_etal08}, we use a simple estimator of the cluster potential based on gas observables. 
Under the assumptions of hydrostatic equilibrium, the gradient of the gravitational potential is related to the 
pressure gradient and density of gas by
\begin{equation}
\nabla \phi = -\frac{\nabla P}{\rho_g},
\label{eq:hse}
\end{equation}
where $\phi$ is the gravitational potential, $\rho_g$ is the density of the intracluster gas, and $P$ is the gas pressure.
The gravitational potential can be obtained as a function 
of gas temperature $T$ and gas density when we integrate Equation~(\ref{eq:hse}) by parts:
\begin{equation}
\phi = -\int \frac{\d P}{\rho_g}+ {\rm constant}=-\frac{k_\B T}{\mu m_\H}-\int\frac{k_\B T}{\mu m_\H}\d \ln\rho_g + {\rm constant}
\end{equation}
where $k_\B$ is the Boltzmann constant, $\mu = 0.59$ is the mean molecular weight for the fully ionized ICM, 
and $m_\H$ is the mass of the hydrogen atom. 

Physical properties relate not to the absolute value of the potential but the difference in the potential. 
The difference of the spherically averaged gravitational potential between two arbitrary radii $r_1$ and $r_2$ can be directly calculated from the temperature and density profiles. We define our potential difference estimator which is based on the temperature and density profiles of the intracluster gas as $\Delta \psi$, 
\begin{equation}
\Delta \psi \equiv \psi(r_2)-\psi(r_1)  = -\left.\frac{kT(r)}{\mu m_\H}\right|^{2}_{1}
                              - \int^{r_2}_{r_1}\frac{kT(r)}{\mu m_\H}\frac{\d\ln\rho_g}{\d r}\d r\,.
\label{eq:pd}
\end{equation}
One can note that $\psi$ is actually the enthalpy of the intracluster gas. If $\Delta \psi$ is a perfect estimator of the true potential difference $\Delta \phi$, then obviously we have $\Delta \psi=\Delta \phi$.  
Since in real clusters, neither the gravitational potential is strictly spherical, nor the intracluster gas obeys perfect hydrostatic equilibrium, we expect  $\Delta \psi$ to deviate from $\Delta \phi$. 
To assess how well $\Delta \psi$ measures $\Delta \phi$ statistically, we fit a scaling relation of the two quantities using
ordinary least squares:
\begin{equation}
\log \Delta \phi = \alpha \log \left(\Delta \psi/\Delta \psi_0\right) + \log \Delta \phi_{0}\,,
\label{eq:fit}
\end{equation}
where we set $\mu m_\H \Delta \psi_0 = 20$ keV. 
Assuming that the residuals follow a log-normal distribution, we calculate the intrinsic scatter 
as the root mean square of the residuals $\delta_{\ln\Delta\phi}$ divided by $\sqrt{N-2}$ where $N$ is the total number of clusters: 
\begin{equation}
\sigma_{\ln\Delta\psi}^2 = \frac{1}{N-2}\sum_i^N \left( \ln\Delta\phi_i-\alpha \ln \left(\Delta \psi/\Delta \psi_0\right) -\ln \Delta \phi_{0}\right)^2.
\end{equation}
Note that we follow the convention of using natural log for the log-normal scatter. 

As $\Delta\psi$ and $\Delta\phi$ are dimensionally the same, we expect the slope of the scaling relation between them to be close to unity. We therefore also compute the scatter of the relation assuming unity slope, i.e., $\alpha=1$. 
In this case, we compute the intrinsic scatter as the root mean square of the residuals divided by $\sqrt{N-1}$ 
because of one less free parameter.  
We estimate the errors of the parameters $\alpha$, $\log \Delta \phi_{0}$ and $\sigma_{\ln\Delta\psi}$
by generating $10^5$ bootstrap samples. 

We calculate the radial profiles for each cluster using logarithmically spaced bins, with a total of 100 bins spanning a 
comoving radial range from $1$ $h^{-1}$~kpc to $5$ $h^{-1}$~Mpc. Note that our results are not affected by the exact details of the binning scheme. The potential difference $\Delta \phi$ for each cluster is calculated from the potential profile directly measured from the simulation. The potential at each radial bin is calculated as the volume-average of the potential over all hydrodynamic cells inside the bin. 

To calculate $\Delta \psi$, we measure the spherically averaged 
gas density profile and temperature profile for each simulated cluster. 
The temperature profile is measured as mass-weighted temperature $T_{\mw}$ which is the
correct representation of the average specific gas thermal energy. For each radial shell,  $T_{\mw}$ is calculated as
\begin{equation}
T_{\mw} = \frac{\sum_i \rho_i T_i \Delta V_i}{\sum_i \rho_i \Delta V_i},
\label{eq:tmw}
\end{equation}
where $\rho_i$, $T_i$, and $\Delta V_i$ are the gas density, gas temperature, and volume of the cell $i$ inside the 
radial shell and summed over all cells inside the radial bin. 

Although mass-weighted temperature can in principle be measured by combining 
X-ray and SZ observations \citep[e.g.,][]{ameglio_etal07,mroczkowski_etal07,puchwein_bartelmann07}, 
X-ray observations alone do not measure the mass-weighted temperature. 
To approximate the temperature measured from X-ray spectra, we calculate $\Delta\psi$ with
the spectroscopic-like temperature $T_{\sl}$ \citep{mazzotta_etal04} which describes the
temperature measured by {\it Chandra} and {\it XMM-Newton} well: 
\begin{equation}
T_{\sl} = \frac{\sum_i \rho_i^2 T_i^{-0.75} T_i\Delta V_i}{\sum_i \rho_i^2 T_i^{-0.75} \Delta V_i}
\label{eq:tsl}
\end{equation}
This weighting scheme preferentially weighs more toward regions with high gas density and low temperature. 
In our calculation of the spectroscopic-like temperature, we exclude cells that have temperature less than $10^6$ K, 
which is well below the instrumental response of current X-ray instruments. 
We denote different $\Delta \psi$ and the fitting parameters that use the mass-weighted temperature and the spectroscopic-like temperature by subscripts ``${\mw}$" and ``${\sl}$", respectively.  

\section{Testing the Estimator}
\label{sec:results}

\begin{table}[htbp]
\begin{center}
\caption{Scaling Relation Parameters at $z=0.0$ for $r_1 = (0.0, 0.2, 0.4)r_{500c}$ and $r_2=r_{500c}$}\label{tab:r500_z0}
\begin{tabular}{ c c | ccc }
\hline
\hline
&&\multicolumn{3}{c}{$z=0.0$} \\
\hline
\multicolumn{2}{c|}{$[r_1/r_{500c},r_2/r_{500c}]=$}&
\multicolumn{1}{c}{$[0.0,1.0]$}&
\multicolumn{1}{c}{$[0.2,1.0]$}&
\multicolumn{1}{c}{$[0.4,1.0]$}
\\
\hline
$\alpha_{\mw}$
&CSF&$0.896\pm 0.098$&$1.018\pm 0.046$&$1.056\pm0.066$\\
&NR  &$0.965\pm 0.030$&$0.970\pm 0.035$&$1.017\pm0.048$\\
$\alpha_{\sl}$
&CSF&$0.984\pm 0.059$&$1.010\pm 0.043$&$1.051\pm0.121$\\
&NR  &$0.951\pm 0.041$&$0.946\pm 0.054$&$0.993\pm0.061$\\
\hline
$\log  (\mu m_\H \Delta \phi_{0,\mw})$
&CSF&$1.607\pm 0.022$&$1.334\pm 0.014$&$1.362\pm0.033$\\
&NR  &$1.327\pm 0.007$&$1.333\pm 0.014$&$1.356\pm0.035$\\
$\log  (\mu m_\H \Delta \phi_{0,\sl})$
&CSF&$1.686\pm 0.013$&$1.375\pm 0.015$&$1.476\pm0.067$\\
&NR  &$1.336\pm 0.008$&$1.342\pm 0.019$&$1.370\pm0.037$\\
\hline
$\sigma_{\ln\Delta\phi,{\mw}}$
&CSF&$0.182\pm 0.027$&$0.088\pm0.021$&$0.140\pm0.035$\\
&NR  &$0.064\pm 0.014$&$0.085\pm0.019$&$0.131\pm0.028$\\
$\sigma_{\ln\Delta\phi,{\sl}}$
&CSF&$0.121\pm 0.017$&$0.072\pm0.016$&$0.190\pm0.024$\\
&NR  &$0.075\pm 0.010$&$0.110\pm0.016$&$0.147\pm0.026$\\
\hline
\hline
&&\multicolumn{3}{c}{Fixed slope $\alpha=1$} \\
\hline
&&\multicolumn{3}{c}{$z=0.0$} \\
\hline
\multicolumn{2}{c|}{$[r_1/r_{500c},r_2/r_{500c}]=$}&
\multicolumn{1}{c}{$[0.0,1.0]$}&
\multicolumn{1}{c}{$[0.2,1.0]$}&
\multicolumn{1}{c}{$[0.4,1.0]$}
\\
\hline
$\log (\mu m_\H \Delta \phi_{0,\mw})$
&CSF&$1.597\pm 0.020$&$1.330\pm 0.009$&$1.335\pm 0.015$\\
&NR&$1.330\pm 0.007$&$1.341\pm 0.009$&$1.347\pm 0.014$\\
$\log (\mu m_\H \Delta \phi_{0,\sl})$
&CSF&$1.686\pm 0.013$&$1.372\pm 0.008$&$1.419\pm 0.019$\\
&NR&$1.340\pm 0.008$&$1.360\pm 0.012$&$1.374\pm 0.015$\\
\hline
$\sigma_{\ln\Delta\phi,{\mw}}$
&CSF&$0.195\pm 0.029$&$0.091\pm 0.023$&$0.148\pm 0.038$\\
&NR&$0.070\pm 0.015$&$0.090\pm 0.018$&$0.136\pm 0.029$\\
$\sigma_{\ln\Delta\phi,{\sl}}$
&CSF&$0.125\pm 0.017$&$0.074\pm 0.017$&$0.198\pm0.020$\\
&NR&$0.083\pm 0.013$&$0.119\pm 0.019$&$0.152\pm0.026$\\
\hline
\end{tabular}
\end{center}
\end{table}

\begin{figure*}[htbp]
\begin{center}
\epsscale{0.75}
\plotone{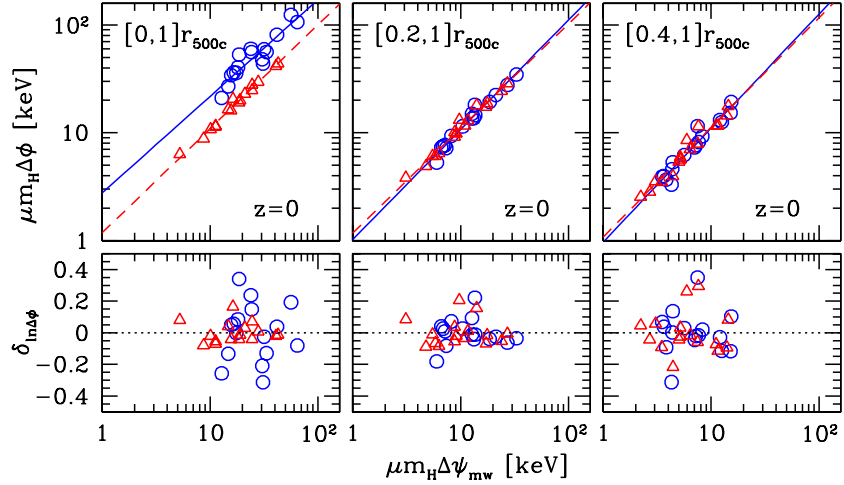}
\caption{ 
Top panels: scaling relations of the potential difference $\Delta \phi$ and an estimator of the potential difference $\Delta \psi$
constructed from gas density and mass-weighted temperature. The potential and the gas observables are measured from three sets of radial separations: 
from the cluster center to $r_{500c}$ ({left} panel), from $0.2r_{500c}$ to $r_{500c}$ ({middle} panel),
and from $0.4r_{500c}$ to $r_{500c}$ ({right} panel).  The blue circles are for the CSF clusters and the
red triangles are for the NR clusters. The blue solid line and the red dashed line are the fitted relation for the CSF and NR
clusters respectively. 
The {bottom} panels show the corresponding residuals of the fits in the top panels. 
}
\label{fig:mw_r500c_z0}
\end{center}
\end{figure*}

We use high-resolution cosmological cluster simulations to compare the potential difference estimator $\Delta \psi$ with the real potential difference $\Delta \phi$ directly measured from the simulation.  We have made following assumptions for our estimator. 
\begin{enumerate}
\item {Input gas physics has little or no effect on the robustness of the estimator. }
\item {The assumption of hydrostatic equilibrium is justified. }
\item {The use of spectroscopic-like temperature is adequate to estimate the potential with little increase in the scatter of the $\Delta\phi$--$\Delta\psi$ relation. }
\item {The $\Delta\phi$--$\Delta\psi$ relation does not evolve with redshift. }
\item {The intrinsic scatter of the $\Delta\phi$--$\Delta\psi$ relation follows log-normal distribution. }
\end{enumerate}

In the following, we test the validity of the assumptions (1) -- (4) and quantify the amount of scatter in the $\Delta\phi$--$\Delta\psi$ relation. For (5), we need a larger sample of clusters to quantify the deviation from log-normal behavior for the scatter which we leave it for future work. 

We report the values of the fitted slopes $\alpha$, normalization $\log (\mu m_\H \Delta \phi_{0})$
and intrinsic scatter $\sigma_{\ln\Delta\phi}$ of the $\Delta \phi$--$\Delta \psi$ scaling relations in upper halves of Table~\ref{tab:r500_z0} -- Table~\ref{tab:csfnr02r500c_zall}. In the lower halves of these tables, we report the normalization and intrinsic scatter where we fit the $\Delta \phi$--$\Delta \psi$ relation with the slope fixed to unity, $\alpha=1$.  We report values of the fitted parameters estimated using both mass-weighted and spectroscopic-like temperatures wherever possible. Since we are measuring and estimating the potential difference between the inner radius $r_1$ and the outer radius $r_2$, for convenience we denote the radial interval for each scaling relation as $[r_1,r_2]$. 

\begin{table}[htbp]
\begin{center}
\caption{Scaling Relation Parameters at $z=0.0,0.6,1.0$ for $r_1=0.2r_{500c}$ and $r_2=r_{500c}$}\label{tab:02r500c_zall}
\begin{tabular}{ c c | ccc }
\hline
\hline
\multicolumn{2}{c|}{$[r_1,r_2] =$}&
\multicolumn{3}{c}{$[0.2r_{500c},r_{500c}]$} \\
\hline
\multicolumn{2}{c|}{$z=$}&
\multicolumn{1}{c}{$0.0$}&
\multicolumn{1}{c}{$0.6$}&
\multicolumn{1}{c}{$1.0$}
\\
\hline
$\alpha_{\mw}$
&CSF&$1.018\pm 0.046$&$0.936\pm 0.050$&$0.912\pm 0.030$\\
&NR  &$0.970\pm 0.035$&$0.901\pm 0.056$&$0.916\pm 0.054$\\
$\alpha_{\sl}$
&CSF&$1.010\pm 0.043$&$0.852\pm 0.070$&$0.895\pm 0.034$\\
&NR  &$0.946\pm 0.054$&$0.875\pm 0.062$&$0.881\pm 0.069$\\
\hline
$\log (\mu m_\H \Delta \phi_{0,\mw})$
&CSF&$1.334\pm 0.014$&$1.324\pm 0.016$&$1.314\pm 0.018$\\
&NR  &$1.333\pm 0.014$&$1.300\pm 0.025$&$1.321\pm 0.027$\\
$\log (\mu m_\H \Delta \phi_{0,\sl})$
&CSF&$1.375\pm 0.015$&$1.350\pm 0.036$&$1.380\pm 0.023$\\
&NR  &$1.342\pm 0.019$&$1.313\pm 0.030$&$1.332\pm 0.037$\\
\hline
$\sigma_{\ln\Delta\phi,{\mw}}$
&CSF&$0.088\pm 0.021$&$0.094\pm 0.018$&$0.069\pm 0.011$\\
&NR  &$0.085\pm 0.019$&$0.150\pm 0.022$&$0.080\pm 0.014$\\
$\sigma_{\ln\Delta\phi,{\sl}}$
&CSF&$0.072\pm 0.016$&$0.150\pm0.022$&$0.076\pm 0.013$\\
&NR  &$0.110\pm 0.016$&$0.130\pm0.014$&$0.107\pm 0.014$\\
\hline
\hline
&&\multicolumn{3}{c}{Fixed slope $\alpha=1$} \\
\hline
\multicolumn{2}{c|}{$[r_1,r_2] =$}&
\multicolumn{3}{c}{$[0.2r_{500c},r_{500c}]$} \\
\hline
\multicolumn{2}{c|}{$z=$}&
\multicolumn{1}{c}{$0.0$}&
\multicolumn{1}{c}{$0.6$}&
\multicolumn{1}{c}{$1.0$}
\\
\hline
$\log (\mu m_\H \Delta \phi_{0,\mw})$
&CSF&$1.330\pm 0.009$&$1.349\pm 0.010$&$1.353\pm 0.009$\\
&NR&$1.341\pm 0.009$&$1.345\pm 0.013$&$1.367\pm 0.010$\\
$\log (\mu m_\H \Delta \phi_{0,\sl})$
&CSF&$1.372\pm 0.008$&$1.418\pm 0.017$&$1.436\pm 0.010$\\
$(\alpha = 1)$
&NR&$1.360\pm 0.012$&$1.372\pm 0.015$&$1.401\pm 0.014$\\
\hline
$\sigma_{\ln\Delta\phi,{\mw}}$
&CSF&$0.091\pm 0.023$&$0.102\pm 0.021$&$0.087\pm 0.011$\\
&NR  &$0.090\pm 0.018$&$0.129\pm 0.020$&$0.099\pm 0.026$\\
$\sigma_{\ln\Delta\phi,{\sl}}$
&CSF&$0.074\pm 0.017$&$0.172\pm0.024$&$0.099\pm 0.013$\\
&NR  &$0.119\pm 0.019$&$0.153\pm0.020$&$0.136\pm 0.035$\\
\hline
\end{tabular}
\end{center}
\end{table}

\begin{table}[htbp]
\begin{center}
\caption{Scaling Relation Parameters at $z=0.0,0.6,1.0$ for $r_1=0.2r_{500c}$ and $r_2=r_{500c}$ for $\Delta \phi^{\rm NR}-\Delta\psi^{\rm CSF}$ Relation }\label{tab:csfnr02r500c_zall}
\begin{tabular}{ c c | ccc }
\hline
\hline
\multicolumn{2}{c|}{$[r_1,r_2] =$}&
\multicolumn{3}{c}{$[0.2r_{500c},r_{500c}]$} \\
\hline
\multicolumn{2}{c|}{$z=$}&
\multicolumn{1}{c}{$0.0$}&
\multicolumn{1}{c}{$0.6$}&
\multicolumn{1}{c}{$1.0$}
\\
\hline
$\alpha_{\mw}$
&&$1.072\pm 0.088$&$1.002\pm 0.074$&$1.024\pm 0.068$\\
$\alpha_{\sl}$
&&$1.070\pm 0.084$&$0.867\pm 0.122$&$0.997\pm 0.086$\\
\hline
$\log (\mu m_\H \Delta \phi_{0,\mw})$
&&$1.271\pm 0.024$&$1.288\pm 0.022$&$1.278\pm 0.036$\\
$\log (\mu m_\H \Delta \phi_{0,\sl})$
&&$1.315\pm 0.026$&$1.295\pm 0.053$&$1.348\pm 0.050$\\
\hline
$\sigma_{\ln\Delta\phi,{\mw}}$
&&$0.184\pm 0.035$&$0.187\pm 0.045$&$0.137\pm 0.020$\\
$\sigma_{\ln\Delta\phi,{\sl}}$
&&$0.164\pm 0.031$&$0.270\pm0.054$&$0.162\pm 0.030$\\
\hline
\hline
&&\multicolumn{3}{c}{Fixed slope $\alpha=1$} \\
\hline
\multicolumn{2}{c|}{$[r_1,r_2] =$}&
\multicolumn{3}{c}{$[0.2r_{500c},r_{500c}]$} \\
\hline
\multicolumn{2}{c|}{$z=$}&
\multicolumn{1}{c}{$0.0$}&
\multicolumn{1}{c}{$0.6$}&
\multicolumn{1}{c}{$1.0$}
\\
\hline
$\log (\mu m_\H \Delta \phi_{0,\mw})$
&&$1.255\pm 0.019$&$1.287\pm 0.019$&$1.267\pm 0.014$\\
$\log (\mu m_\H \Delta \phi_{0,\sl})$
&&$1.297\pm 0.017$&$1.356\pm 0.028$&$1.350\pm 0.017$\\
\hline
$\sigma_{\ln\Delta\phi,{\mw}}$
&&$0.182\pm 0.036$&$0.180\pm 0.042$&$0.133\pm 0.029$\\
$\sigma_{\ln\Delta\phi,{\sl}}$
&&$0.163\pm 0.031$&$0.269\pm 0.058$&$0.157\pm 0.028$\\
\hline
\end{tabular}
\end{center}
\end{table}

\subsection{Effects of Dissipation}
\label{subsec:gphysics}
 
Assuming hydrostatic equilibrium, we measure $\Delta \phi$ and $\Delta \psi_{\mw}$ 
in the simulated clusters for both CSF and NR clusters at $z=0$ for different radial distance
separations: $[r_1,r_2]$ where we set $r_1/r_{500c} = 0, 0.2$ and $0.4$, and fix $r_2 = r_{500c}$,  
as it is the typical outermost radius current X-ray surveys are capable of
measuring. This is also the radius within which the cluster is relaxed \citep{evrard_etal96} 
and where the non-thermal pressure small, $\lesssim 10\%$  \citep[e.g.,][]{evrard90,rasia_etal06,nagai_etal07b}. 
However, as we show later the estimator is robust to the exact choice of radial separation. 

In Figure~\ref{fig:mw_r500c_z0}, we plot the $\Delta \phi$--$\Delta \psi_{\mw}$ relations 
for $[r_1,r_2]/r_{500c} = [0,1],[0.2,1],[0.4,1]$ for both CSF and NR clusters. 
All slopes of the relations are within the expected value of unity to within $1\sigma$.  
The normalization $\log(\mu m_\H \Delta \psi_0)$ is also near to the expected value of $\log 20 \approx 1.3$. 
The intrinsic scatter varies from $\sim 6\%$ to $15\%$, depending on the radial separation and input cluster physics. 
In general the resulted scatters of the fixed slope relations increase slightly compared to those where the slope is a free parameter. 
The normalizations of the fixed slope relation remain unchanged to within $1\sigma$. The values of the fitted parameters are reported in Table~\ref{tab:r500_z0}. 

For $[r_1,r_2]/r_{500c} = [0.2,1],[0.4,1]$, the slope, normalization and scatter are similar for both CSF and NR clusters. 
Dissipative gas physics has little effect on the potential estimator outside the cluster core $r_1 \gtrsim 0.2r_{500c}$. 

\begin{figure}[htbp]
\begin{center}
\epsscale{1.22}
\plotone{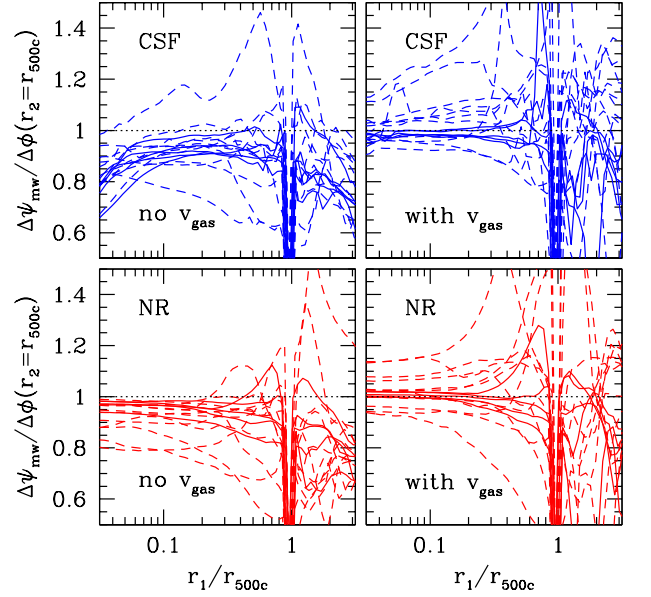}
\caption{$\Delta \psi_{\mw}/\Delta \phi$ as a function of the inner radius $r_1$ with the outer radius $r_2 = r_{500c}$ at $z=0$. 
The solid lines represent relaxed clusters and the dashed lines represent unrelaxed clusters. 
Top panels show the CSF clusters and the bottom panels show the NR clusters. The left panels show the fiducial $\Delta \psi_{\mw}$ without including gas motions, while the right panels include pressure support form gas motions. 
}
\label{fig:dpsi_r2r500c}
\end{center}
\end{figure}

For $[r_1,r_2]/r_{500c} = [0,1]$, while the slope is similar in both CSF and NR clusters, the normalization of the CSF relation
is offset to a higher value compared to the NR relation. As we discuss in Section~\ref{subsec:hse}, the higher normalization is due to strong gas rotation in the CSF clusters, where $\Delta \psi$ underestimates $\Delta \phi$, shifting the data points to the left in the 
$\Delta \phi$--$\Delta \psi_{\mw}$ plane. This underestimation is also shown in the {upper left} panel of Figure~\ref{fig:dpsi_r2r500c}, 
where $\Delta \psi_{\mw}/\Delta \phi$ is plotted as a function of the inner radius $r_1$.  The CSF relation have a relatively large scatter of about $18\%$ due to strong dissipation in the cluster core regions. 

The potential estimator  $\Delta \psi$ is robust to dissipational gas physics outside the cluster core. As we show in Section~\ref{subsec:hse} that once we correct for non-thermal pressure support due the gas motions, the estimator becomes robust to gas physics even when the cluster core is included. 

\begin{figure}[t]
\begin{center}
\epsscale{1.2}
\plotone{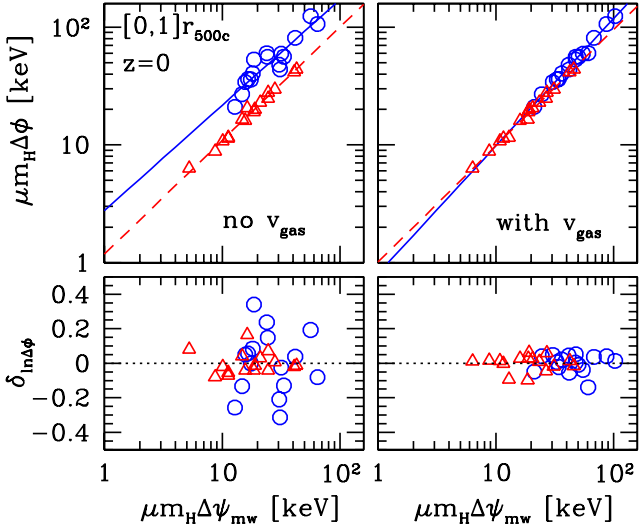}
\caption{Top panels: Scaling relations of the potential difference $\Delta \phi$ and an estimator of the potential difference $\Delta \psi$
constructed from gas density and mass-weighted temperature. The potential and the gas observables are measured from the cluster center to $r_{500c}$. The left panel shows the scaling relation without including pressure due to gas motions, while right panel shows the scaling relation with gas motions. 
The blue circles are for the CSF clusters and the red triangles are for the NR clusters. The blue solid line and the red dashed line are the fitted relation for the CSF and NR clusters respectively. 
The bottom panels show the corresponding residuals of the fits in the top panels. 
}
\label{fig:gv_0r500c_z0}
\end{center}
\end{figure}

\subsection{Non-thermal Pressure Support}
\label{subsec:hse}

\begin{figure*}[htbp]
\begin{center}
\epsscale{0.75}
\plotone{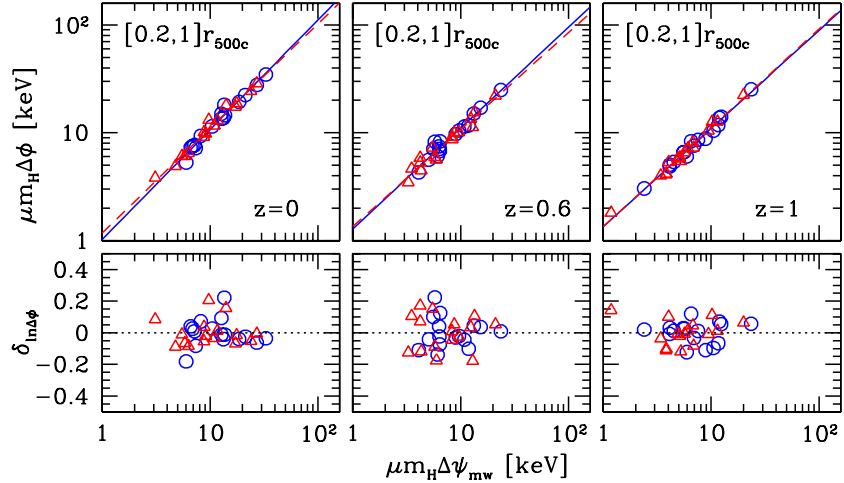}
\caption{Top panels: scaling relations of the potential difference $\Delta \phi$ and
the potential difference estimator $\Delta \psi_{\mw}$
for a fixed radial separation with the inner radius $r_1= 0.2r_{500c}$ and the outer radius $r_2=r_{500c}$ 
at different redshift: $z=0$ (left panel), $z=0.6$ (middle panel), and $z=1$ (right panel).  
The blue circles are for the CSF clusters and the red triangles are for the NR clusters. 
The blue solid line and the red dashed line are the fitted relation for the CSF and NR clusters 
respectively. The bottom panels show the corresponding residuals of the fits in the top panels. }
\label{fig:mw_02r500c_zall}
\end{center}
\end{figure*}

We test the validity of the assumption of hydrostatic equilibrium of our potential estimator by explicitly including the contribution of non-thermal pressure support by gas motions, which are mainly responsible for the departure of hydrostatic equilibrium in our
simulated clusters \citep{lau_etal09}. The effects of gas motions can be corrected by including the following
term $\Delta{\psi}_{\rm v_{gas}}$ in Equation~(\ref{eq:pd}), derived from the spherically symmetric radial Jean's equation  
\citep[e.g.,][]{binney_tremaine08}:
\begin{equation}
\Delta{\psi}_{\rm v_{gas}} = 
-\left(\overline{v_r^2}(r_2)-\overline{v_r^2}(r_1)\right)
-\int^{r_2}_{r_1}\overline{v_r^2}\frac{\d\ln\rho_g}{\d r}{\d r}  
-\int^{r_2}_{r_1}\frac{ 2\overline{v_r^2}-\overline{v_t^2} }{r}{\d r}        
\label{eq:pdvg}
\end{equation}
where $\overline{v_r^2}$ and $\overline{v_t^2}$ are the radial and tangential components of the
mean-square gas velocities averaged over the spherical radial shell, with respect to the cluster peculiar velocity
defined as the average mass-weighted velocity of DM inside $r_{500c}$. 

Figure~\ref{fig:gv_0r500c_z0} shows the scaling relation with and without the inclusion of
the non-thermal pressure term $\Delta{\psi}_{\rm v_{gas}}$. 
Including gas motions, i.e., replacing $\Delta{\psi}_{\mw}$ with $\Delta{\psi}_{\mw}+ \Delta{\psi}_{\rm v_{gas}}$,
results in better estimates of the potential difference
by decreasing the scatter of the scaling relation to $\sim 5\%$ for $[0,1]r_{500c}$ for both NR and CSF clusters
This is illustrated more clearly in Figure~\ref{fig:dpsi_r2r500c} where 
we plot the deviations of the estimated potential difference $\Delta \psi_{\mw}$ 
from the true potential difference $\Delta \phi$ as a function of $r_1$ with $r_2$  fixed at $r_{500c}$, with
solid lines representing the dynamically relaxed clusters and dashed lines representing the unrelaxed clusters. 
Without the inclusion of gas motions, $\Delta \psi_{\mw}$ underestimates $\Delta \phi$ 
by about $\sim 5\%$--$20\%$ for $r_1\gtrsim 0.1r_{500c}$ in both NR and CSF clusters, a value that is 
consistent with gas motions biasing the thermal pressure low by about the same amount \citep{lau_etal09}. 
Most of the scatter is driven by the dynamically disturbed systems in the sample. 
For $r_1$ approaching $r_2=r_{500c}$, the scatter in both CSF and NR clusters increases as 
the outer regions of the cluster are less relaxed. 
In the inner region $r_1 \lesssim 0.1r_{500c}$, the deviations from hydrostatic equilibrium
are approximately constant for the NR clusters, but for CSF clusters, 
$\Delta \psi_{\mw}$ increasingly underestimates $\Delta \phi$ as $r_1$ 
decreases due to strong gas rotation near the core where the gas is rotationally supported induced by gas cooling 
\citep{lau_etal10}. Including gas motions takes into the account the effective potential due to the rotation,
leading to a better recovery of the true potential difference and agreement between the scaling relations of the CSF and the NR clusters. This is shown in the upper-right panel of Figure~\ref{fig:gv_0r500c_z0}. 

\subsection{Spectroscopic-like Temperature}

The spectroscopic-like temperature $T_{\sl}$ (Equation~(\ref{eq:tsl})) 
is generally less than the mass-weighted temperature $T_{\mw}$ (Equation~(\ref{eq:tmw})), 
as it weighs toward colder and denser gas. As shown in Table~\ref{tab:r500_z0},
for different $[r_1,r_{500c}]$ the scatter in the $\Delta\phi$--$\Delta\psi_{\sl}$ relation
is also larger than  $\Delta\phi$--$\Delta\psi_{\mw}$ relation, as $\Delta\psi_{\sl}$ is biased towards
the colder and denser gas, whose fractions and locations vary across different clusters. 
Nevertheless, the scatter is increased only by a few percent, with the exception 
for $[r_1,r_2] = [0.4,1]r_{500c}$ where the large scatter is driven by a single cluster which has
abnormally low $T_{\sl}$ due to a dense gas clump. 

Comparing the $\Delta\phi$--$\Delta\psi_{\sl}$ relation between the CSF and NR clusters,  we find that 
the normalization for the CSF relations are slightly higher than their NR counterparts at large $r_1$, 
because of the relatively larger number of cold dense clumps in the CSF clusters
than in the NR clusters. Nevertheless their normalization agree to within 1$\sigma$. 

To account for the limited angular resolution of X-ray observations, when we measure $T_{\sl}$,  
we average it over a radial window of $\Delta r =100$~kpc centered on the radius of interest. 
Varying the size of this window from $50$~kpc to $200$~kpc has little effect in changing 
the $\Delta\phi$--$\Delta\psi_{\sl}$ relations. 
 
 \subsection{Evolution with Redshift}
\label{subsec:z}
 
Next, we investigate the evolution of the $\Delta \phi$--$\Delta \psi$ relations by fixing 
$[r_1,r_2] = [0.2,1]r_{500c}$ and fit the relations at $z=0.0,0.6$ and $1.0$.  
Figure~\ref{fig:mw_02r500c_zall} shows the resulting plots. All the relations at higher redshift are
consistent with the $z=0$ results to within $2\sigma$. Scatter decreases slightly 
with increasing redshift and there is a weak trend of decreasing slope with redshift. 
The values of the parameters are shown in Table~\ref{tab:02r500c_zall}. 

The apparent weak evolution in the  $\Delta \phi$--$\Delta \psi$  relation 
can perhaps be explained by the lack of change in the gravitational potential over time once
the DM halo has formed.  For example, \cite{li_etal07} demonstrated semi-analytically 
that the circular velocity at the virial radius of the halo remains relatively unchanged throughout much of its
mass accretion history. Given that our sample size is quite small (16 clusters), 
the apparent redshift evolution could be driven by a few clusters. A complete understanding of the redshift evolution of the potential well and a full characterization of the redshift evolution of the $\Delta \phi$--$\Delta \psi$  relation will require tests using 
simulations with more clusters and redshift outputs than the current study. 

\subsection{Choice of Radial Separations}
\label{subsec:radial}

\begin{figure*}[htbp]
\begin{center}
\epsscale{0.75}
\plotone{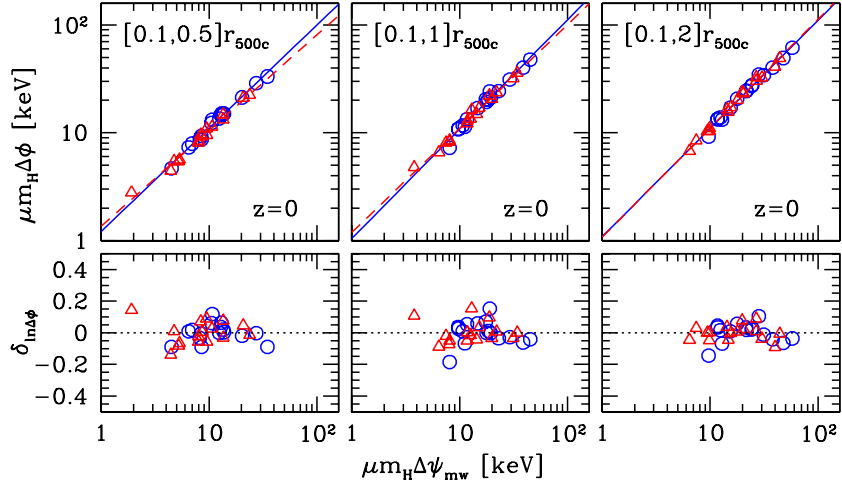}
\caption{Top panels: scaling relations of the potential difference $\Delta \phi$ and
the potential difference estimator $\Delta \psi_{\mw}$ at $z=0$
for inner radius $r_1= 0.1r_{500c}$ and outer radius $r_2=0.5r_{500c}$ (left panel), 
$r_2=r_{500c}$ (middle panel) and $r_2=2r_{500c}$ (right panel).  
The blue circles are for the CSF clusters and the red triangles are for the NR clusters. 
The blue solid line and the red dashed line are the fitted relation for the CSF and NR clusters 
respectively. The bottom panels show the corresponding residuals of the fits in the top panels. }
\label{fig:mw_01r500c_z0}
\end{center}
\end{figure*}

In principle $r_1$ and $r_2$ that define $\Delta \phi \equiv \phi(r_2)-\phi(r_1)$ and $\Delta \psi$  
can be chosen arbitrary. In practice, the best choice of $[r_1,r_2]$  would be the one 
that gives the lowest scatter. Once corrected for pressure from gas motions, the larger $|r_1-r_2|$ will result in less scatter,
as the integral in $\Delta \psi$ (Equation~(\ref{eq:pd})) encompasses larger radial range thus
making the estimator more robust to small-scale variations in gas density and temperature. 
For example, setting $[r_1,r_2] =[0.1,2]r_{500c}$ gives a scatter of about $\sim 6\%$, compared to
$\sim 8\%$ for $[r_1,r_2] =[0.1,1]r_{500c}$. 
Decreasing $r_1$ also gives smaller scatter as long as one stays away from the cluster core for the CSF clusters. 
For example,the scatter decreases from $\sim 14\%$ for $[r_1,r_2] =[0.4,1]r_{500c}$ to $\sim 9\%$ for $[r_1,r_2] =[0.2,1]r_{500c}$. 
For $r_2 > r_{500c}$, the cluster gas deviates from hydrostatic equilibrium significantly with radius 
and  $\Delta \psi$ underestimates the true potential difference  $\Delta \phi$. However, this underestimation is small.
As shown in Figure~\ref{fig:mw_01r500c_z0}, the slope and normalization of the scaling relation are essentially the same 
as the radial interval changes from  $[r_1,r_2] =[0.5,1]r_{500c}$ to $[r_1,r_2] =[0.1,2]r_{500c}$. 
The underestimation in the potential is significantly less than that of the hydrostatic mass estimate because $\Delta \psi$ encompasses a large region of the cluster interior where deviation from hydrostatic equilibrium is small, while the hydrostatic cluster mass measured at a particular radius is dependent on the local value of the pressure gradient. 

\begin{figure*}[htbp]
\begin{center}
\epsscale{0.75}
\plotone{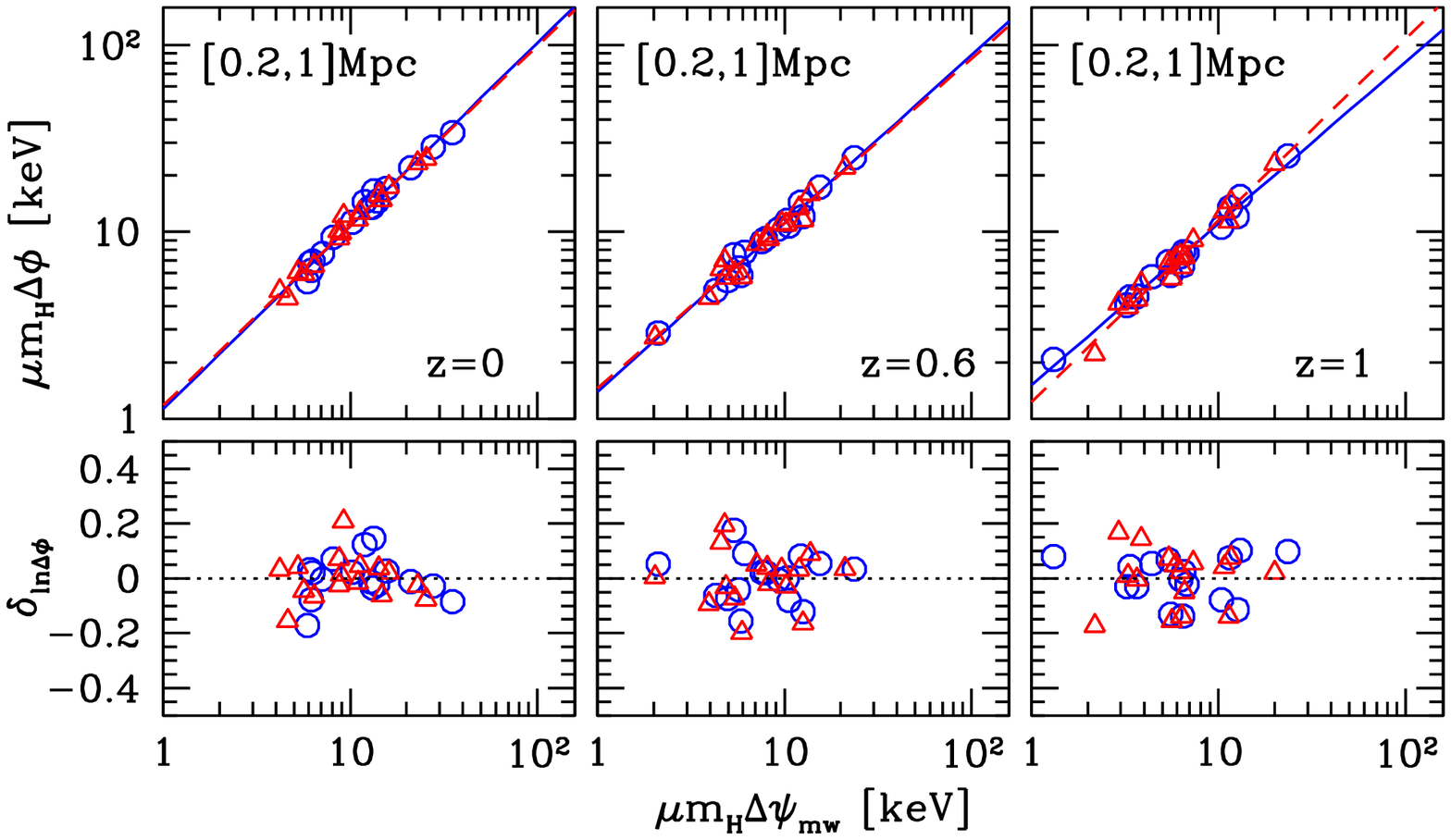}
\caption{Top panels: scaling relations of the potential difference $\Delta \phi$ and
the potential difference estimator $\Delta \psi_{\mw}$
for a fixed radial separation with the inner radius $r_1= 0.2$ Mpc and the outer radius $r_2=1$ Mpc 
at different redshift: $z=0$ (left panel), $z=0.6$ (middle panel), and $z=1$ (right panel).  
The blue circles are for the CSF clusters and the red triangles are for the NR clusters. 
The blue solid line and the red dashed line are the fitted relation for the CSF and NR clusters 
respectively. The bottom panels show the corresponding residuals of the fits in the top panels. }
\label{fig:mw_02mpc_zall}
\end{center}
\end{figure*}

Following similar logic, the choice of $r_1$ and $r_2$ should not be limited to functions
of $r_{500c}$. Defining $r_1$ and $r_2$ through $r_{500c}$ can be undesirable as $r_{500c}$
depends on cosmology through the Hubble parameter $H(z)$. 
One choice of the radial interval, independent of the halo radius, is simply the physical distance separations.  
Figure~\ref{fig:mw_02mpc_zall} shows the scaling relations for $[r_1,r_2]= [0.2,1]$Mpc at $z=0.0$, $0.6$ 
and $1.0$. For this radial separation the $\Delta \phi$--$\Delta \psi$ relation has a relatively small scatter of $\sim$ $8\%$--$9\%$ 
for the given redshifts. The values of slope and the normalization of the relations are similar to 
those with $r_1$ and $r_2$ being functions of $r_{500c}$ . The relation is the same 
for the different input gas physics. Redshift evolution of the relation is weak but requires more
investigation as discussed in Section~\ref{subsec:z}.  

\subsection{Calibration for $N$-body Simulations}
\label{subsec:nbody}

\begin{figure*}[htbp]
\begin{center}
\epsscale{0.75}
\plotone{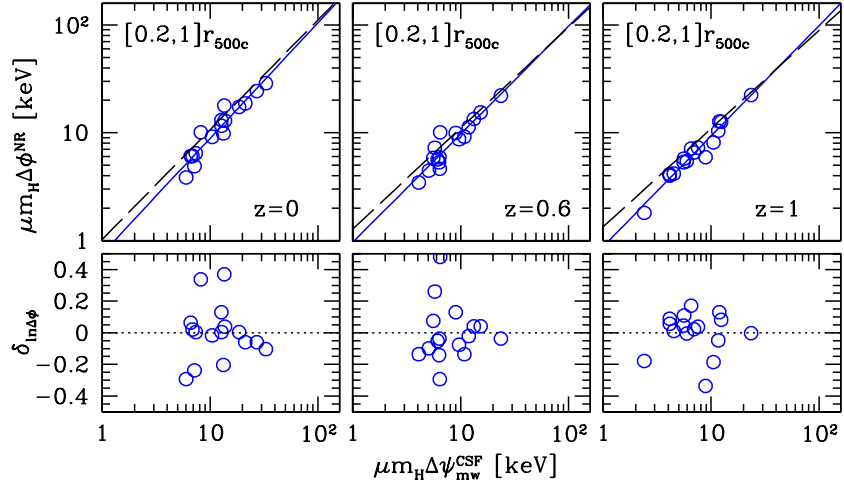}
\caption{Top panels: scaling relations of the potential difference $\Delta \phi^{\rm NR}$ in the NR clusters and
the potential difference estimator $\Delta \psi_{\mw}^{\rm CSF}$ of the corresponding CSF clusters,
for a fixed radial separation with the inner radius $r_1= 0.2r_{500c}$ and the outer radius $r_2=0.5r_{500c}$ 
at different redshift: $z=0$ (left panel), $z=0.6$ (middle panel) and $z=1$ (right panel).  
The blue circles are for the CSF clusters and the red triangles are for the NR clusters. 
The blue solid line is fitted relation. The black dashed line is the best fit for the 
$\Delta \phi^{\rm CSF}$--$\Delta \psi_{\mw}^{\rm CSF}$ relation. 
The bottom panels show the corresponding residuals of the fits in the top panels. }
\label{fig:mwcsf_02r500c_zall}
\end{center}
\end{figure*}

Although becoming increasingly feasible, full cosmological hydrodynamical simulations of cluster formation 
with realistic physics like radiative cooling, star formation and active galactic nucleus feedback are expensive to run. 
A cheaper way would be using these simulations to calibrate less expensive
dissipationless $N$-body simulations. 
Since both pure $N$-body simulations and our NR clusters have gravity as the 
only driving physics, the results of our NR clusters should be similar to the
that of the $N$-body results. 
In this subsection, we therefore investigate how well our 
more realistic CSF clusters measure the gravitational potential wells of the NR clusters, and
therefore the potential wells of clusters in $N$-body simulations.  
The results will be useful for calibrating a cluster potential function from 
pure $N$-body simulations. 

We estimate the NR potential difference $\Delta\phi^{\rm NR}$ using the potential estimator of the CSF
clusters $\Delta\psi^{\rm CSF}_{\mw}$. In Figure~\ref{fig:mwcsf_02r500c_zall}, we show the comparison 
between the $\Delta\phi^{\rm NR}$--$\Delta\psi^{\rm CSF}_{\mw}$ relation for $[r_1,r_2] =[0.2,1]r_{500c}$ at $z=0.0,0.6$, and $1.0$ 
(in blue solid lines) and the $\Delta\phi^{\rm CSF}$--$\Delta\psi^{\rm CSF}_{\mw}$ relation (in black dashed lines). 
The slope of the $\Delta\phi^{\rm NR}$--$\Delta\psi^{\rm CSF}_{\mw}$ relation is steeper than the
$\Delta\phi^{\rm CSF}$--$\Delta\psi^{\rm CSF}_{\mw}$ relation. This is because dissipation results in 
deeper potential wells in the CSF clusters compared to their NR counterparts, and this effect 
is stronger in small size halos which have lower virial temperature. Therefore $\Delta\psi^{\rm CSF}_{\mw}$ 
overestimates the potential difference  $\Delta\phi^{\rm NR}$ more for the smaller size systems, leading to 
a steeper slope in the $\Delta\phi^{\rm NR}-\Delta\psi^{\rm CSF}_{\mw}$ relation. 
The scatter in $\Delta\phi^{\rm NR}$--$\Delta\psi^{\rm CSF}_{\mw}$ relation also increases because the NR clusters are 
not in exact identical dynamical states as their CSF counterparts as 
the dynamical evolution of each cluster halo is affected by the dissipative gas physics which changes 
the structure and hence dynamics of the halos. Table~\ref{tab:csfnr02r500c_zall} summarizes the fitted parameters.

\section{Summary and Discussion}
\label{sec:summary}

\subsection{Summary of Key Results}

In this paper we propose a simple estimator for the gravitational potential difference in clusters of galaxies. 
This estimator is based on the density and temperature profiles of the intracluster gas under the
assumptions of hydrostatic equilibrium and spherical symmetry. 

Using high resolution cosmological hydrodynamical simulations of galaxy clusters, 
we have tested that this estimator is a robust and accurate estimator of the true gravitational potential
of simulated clusters. The key results are summarized below:
\begin{enumerate}
\item {The scaling relation between this estimator and the true gravitational potential difference $\Delta\phi$--$\Delta\psi$ has an intrinsic scatter of $\sim$ $8\%$--$15\%$. }
\item {Input gas physics has little effect on the $\Delta\phi$--$\Delta\psi$ relation as long as the cluster core is excised. }
\item {The assumption of hydrostatic equilibrium is justified outside the cluster core. Adding non-thermal pressure support reduces the scatter in the $\Delta\phi$--$\Delta\psi$ relation for all radii.  The core need not be excised when non-thermal pressure support is included. }
\item {With the core excised or kinetic pressure from gas motions included, the slope and normalization of the $\Delta\phi$--$\Delta\psi$ scaling relation are relatively independent of the choice of radial separations. The scatter of the relation decreases as the separation increases. }
\item {The use of spectroscopic-like temperature is adequate to estimate the potential with little increase in scatter in the $\Delta\phi$--$\Delta\psi$ relation. }
\item {The $\Delta\phi$--$\Delta\psi$ relation evolves weakly with redshift. A full characterization of redshift evolution is needed. }
\end{enumerate}

The results presented above suggest
that gravitational potential can serve as a useful alternative to cluster mass for cluster cosmology. 
X-ray mass-observable scaling relations are usually dependent on baryonic physics. For example, the normalization of
the current tightest mass-observable relation, the $M$--$Y_{\rm X}$ relation \citep{kravtsov_etal06}, 
is dependent on non-thermal pressure support; the normalization of the cluster mass--gas mass $M$--$M_{\rm gas}$ relation, is affected by the input gas physics \citep{fabjan_etal11}. Our results show that the potential scaling relation, with appropriate radial range selection, is little affected by both of these effects. 

Gravitational potential has another advantage over mass in that it is more spherical than that of the matter distribution.\footnote{We give a comparison of the shape of the dark matter distribution and the shape of the gravitational potential well of the same set of clusters used in this paper in \citet{lau_etal10}. } The assumption of spherical symmetry works better for gravitational potential than matter density, and this is perhaps partly why the potential scaling relation has relatively small scatter compared to mass-observable relations. 

An essential characteristic of the gravitational potential scaling relation is that we are free to choose 
the radius where we measure the potential, not limiting to the halo radius defined in terms of overdensity with respect to the mean density of critical density of the universe, both of them functions of cosmology. This takes away the rather arbitrary nature of cluster mass, where different cluster mass definitions give different cluster mass functions \citep[e.g.,][]{white02, tinker_etal08}. 

Despite the advantages given above, there are also several caveats and uncertainties 
in using the proposed potential estimator for cluster cosmology. 
We see a weakly evolving redshift evolution in the potential scaling relation that needs to be 
understood well before using gravitational potential for cluster cosmology. 
Although the potential measured at different radial separations result in essentially the same
potential scaling relation, it is uncertain whether the different radial separation 
would result in different cluster potential function. 
Furthermore, we have assumed spherical symmetry for the gravitational potential and the gas distribution, 
but have not tested for the effect of deviation from spherical symmetry, although we expect
that the effect to be small compared to the effect on cluster mass. 
Since our gravitational potential estimator involves an integral over radius of
gas temperature and density, which is the dominant term in the estimator (Equation~(\ref{eq:pd})), 
the gas density and temperature profiles may need to be measured with high spatial resolution. 
Accurate measurements of gas temperature also require high-resolution X-ray spectrometers. 
This can be technically challenging, especially if we want to measure the potential of high-$z$ clusters.  
However, we note that combined X-ray/SZ analysis is able to recover the temperature profile 
without the need of X-ray spectra \citep[e.g.,][]{nord_etal09}. 

\subsection{Toward the Cluster Potential Function}

Given the results presented in the paper, it will be interesting to see whether the cluster potential function 
is indeed a better alternative for constraining cosmology than the commonly adopted cluster mass function. 
In this subsection, we discuss ways to improve the potential scaling relations and to construct the cluster potential function. 

We suggest that future modeling efforts toward the goal of using the cluster potential function for cluster cosmology 
should focus on two main areas:
(1) further investigation on the intrinsic nature of the potential scaling relation and the properties of the cluster potential function and
(2) implementation of the potential scaling relation in actual observations. 

The first aspect aims to answer the question whether the cluster potential function can replace the cluster mass function 
for cosmology. To see whether the cluster potential function has more
constraining power than the cluster mass function, comparison between the cluster potential function 
and the cluster mass function calibrated from the same large-scale simulation is required. The 
comparison will have to focus on whether cluster function recovers the fiducial cosmological parameters better in terms of 
degeneracies between parameters, statistical errors, and systematic uncertainties. 
Simulations with statistically large sample of clusters
at multiple redshifts will help constraining the weak redshift evolution of the potential scaling relation shown in the current work. 
They will also help further characterize the scatter and systematics of the potential scaling relation, for example, deviations
from lognormal distribution in the scatter.

The second aspect focuses on developing new ways of characterizing gravitational potentials based on cluster observables
and apply them to synthetic and real observations. For example, the potential estimator proposed in the 
current work will benefit from mock X-ray analyses in addressing systematics like gas clumping \citep{mathiesen_etal99,simionescu_etal11,nagai_lau11}
and some of its observational challenges discussed in the previous subsection. 
It is reasonable to expect that other types of cluster potential estimators
exist and may even perform better than the one proposed in this paper. For example, gravitational lensing signals like shear and convergence depend on the projected gravitational potential and may provide a new probe to the gravitational potential.  Velocity dispersion of cluster galaxies directly probes the cluster potential, although it can be affected by velocity anisotropy and projection effects. SZ signals measure the integrated gas pressure, and it is similar to the gas enthalpy-based estimator proposed in the current work. Multiple probes of the gravitational field will be helpful in determining systematics, and the covariance of different potential estimator may also improve the precision of the measured potential, as it is for the case of cluster mass proxies \citep{stanek_etal10}. 

Clearly, much more work is needed to address these issues and we hope that this paper provide a starting point for the investigation  of using the cluster potential function for cosmology. 

\acknowledgements
The author thanks the anonymous referee whose suggestions greatly improve the content and presentation 
of this paper. The author also thanks his PhD advisor Andrey Kravtsov for his continuous support,
guidance and patience, and Daisuke Nagai for kindly providing the simulation data and offering invaluable advice. 
The author also thanks Mike Gladders, Nick Gnedin, and Rick Kron for helpful comments. 
The author is supported by the NSF grant AST-0708154 and by NASA grant NAG5-13274. 
The cosmological simulations used in this study were performed on the IBM RS/6000 SP4 system (copper) at the
National Center for Supercomputing Applications (NCSA).  
This work made extensive use of the NASA Astrophysics Data System and arXiv.org preprint server.
\\

\bibliography{ms}

\end{document}